\documentclass[aps,prl,twocolumn,amsmath,amssymb,floatfix,
superscriptaddress]{revtex4}

\usepackage{graphicx}

\def\bra#1{\left\langle#1\right|}
\def\ket#1{\left|#1\right\rangle}

\def\kd#1{\left[#1\right]}

\def\be{\begin{equation}}       \def\ee{\end{equation}}
\def\bea{\begin{eqnarray}}      \def\eea{\end{eqnarray}}
\def\ba{\begin{array} }
\def\ea{\end{array} }
\def\bnum{\begin{enumerate} }
\def\enum{\end{enumerate}}

\def\nn{\nonumber}

\def\=>{\Rightarrow}
\def\>{\rightarrow}

\def\eye2{Fathbb{I}}

\def\te{\mathrm{e}}

\def\Tr{\mathrm{Tr}}

\begin{document}
\title{Entanglement entropy and entanglement spectrum of the Kitaev model}
\author{Hong Yao}
\affiliation{Department of Physics, University of California,
Berkeley, CA 94720, USA} \affiliation{Materials Sciences Division,
Lawrence Berkeley National Laboratory, Berkeley, CA 94720, USA}
\author{Xiao-Liang Qi}
 \affiliation{Microsoft Research, Station Q, Elings
Hall, University of California, Santa Barbara, CA 93106, USA}
\affiliation{Department of Physics, Stanford University, Stanford,
CA 94305, USA}
\date{\today}
\begin{abstract}
In this paper, we obtain an exact formula for the entanglement
entropy of the ground state and all excited states of the Kitaev
model. Remarkably, the entanglement entropy can be expressed in a
simple separable form $S=S_G+S_F$, with $S_F$ the entanglement
entropy of a free Majorana fermion system and $S_G$ that of a $Z_2$
gauge field. The $Z_2$ gauge field part contributes to the universal
``topological entanglement entropy'' of the ground state while the
fermion part is responsible for the non-local entanglement carried
by the $Z_2$ vortices (visons) in the non-Abelian phase. Our result also enables the calculation of the entire entanglement spectrum and
 the more general Renyi entropy of the Kitaev model. Based on our results we propose a new quantity to characterize topologically ordered states---the {\em
 capacity of entanglement}, which can distinguish the states
 with and without topologically protected gapless entanglement spectrum.
\end{abstract}
\maketitle

Exotic phases such as fractional quantum Hall (FQH) states, which are not in the paradigm of conventional symmetry breaking, were termed as topologically ordered \cite{Wen90}
since they have robust ground state degeneracy which is protected
against all local perturbations, but sensitive to the topology of
the system
\cite{Kivelson87}.
A topologically ordered state has non-local pattern of quantum entanglement, which is essential for the proposal of topological quantum computation \cite{Kitaev03,dassarma2005,nayak2008}.

By bipartitioning a system spatially, the entanglement entropy (EE)
measures how closely entangled the two subsystems are. For a gapped
system, EE is usually proportional to the area of the interface
between the two subsystems in the thermodynamic limit. However, as
discovered \cite{LevinWen,KitaevPreskill} by Levin and Wen as well as Kitaev and
Preskill, the entanglement entropy of a
topologically ordered state contains a universal constant term, which
is uniquely determined by the topological order of the state, named
as topological entanglement entropy (TEE). TEE enables a direct
characterization of topological ordered states without referring to
the Hamiltonian. EE and TEE are properties of a many-body state and
are usually hard to compute. EE and/or TEE have been computed exactly or numerically for several models such as toric code model\cite{Kitaev03,hamma2005,Chamon07}, FQH states\cite{Laughlin83,Haque07,Zuzolia07} and quantum dimer models\cite{Rokhsar88,Fradkin06,Furukawa07}. So far there has been no exact result for the EE
of topologically ordered states whose quasiparticles obey non-Abelian statistics.

This paper serves to fill in that gap by providing a simple but
exact method to compute the EE for any eigenstate (either ground or
excited states) of the Kitaev model \cite{Kitaev06}, which is one of
the most important exact solvable models with non-Abelian
anyons.
The essence of our method is a rigorous proof that the EE of the
Kitaev model is equal to that of two decoupled systems: a sourceless
$Z_2$ gauge field and a free Majorana fermion system. Although the
TEE of the ground state comes only from the $Z_2$ gauge field, the
fermionic part is responsible for all nontrivial entanglement
properties of the non-Abelian phase. Besides the EE, our method also
enables the computation of the whole entanglement spectrum (ES), {\it
i.e.}, the eigenvalue spectrum of the reduced density
matrix \cite{Li07}. We show that the entanglement spectrum is gapless
or gapped in the non-Abelian and Abelian phase of the Kitaev model,
respectively. We propose a new quantity, the {\it capacity of entanglement},
which can be used to distinguish different topological states with
gapped and gapless entanglement spectrum.

Kitaev model is a spin-1/2 model originally proposed on the honeycomb
lattice \cite{Kitaev06} with the Hamiltonian
\begin{eqnarray}
H=-\sum_{x\text{-link}}J_x\sigma^x_i\sigma^x_j -\sum_{y\text{-link}}J_y\sigma^y_i\sigma^y_j -\sum_{z\text{-link}}J_z\sigma^z_i\sigma^z_j,
\end{eqnarray}
where $x,y,z$-link stand for the three types of links.
 It has a non-Abelian phase when the
time-reversal symmetry is broken either explicitly by magnetic field
\cite{Kitaev06} or three-spin couplings \cite{Lee07}, or
spontaneously by decorating the honeycomb lattice
\cite{Yao07}. For simplicity, hereafter we will present our
exact results of
EE and entanglement spectrum for
the Kitaev model on honeycomb lattice, but our approach can be
generalized straightforwardly to a broad class of Hamiltonians,
including the Kitaev model on any trivalent lattice and Gamma matrix
models \cite{Wen03,Yao09,Wu09,Ryu09,nussinov2009}.

The Kitaev model can be solved by introducing the Majorana
representation of the Pauli matrices \cite{Kitaev06}:
$\sigma^\alpha_i=i\gamma^\alpha_i \eta_i$ ($\alpha=x,y,z$), where
$\gamma^\alpha_i$ and $\eta_i$ are Majorana fermion operators.
$\gamma^\alpha_i$ and $\eta_i$ on each lattice site define a
$4$-dimensional Hilbert space, so that the Majorana representation
of a spin-$1/2$ is redundant. The physical Hilbert space is defined
by a constraint $D_i=-i\sigma^x_i\sigma^y_i\sigma^z_i
=\gamma^x_i\gamma^y_i\gamma^z_i\eta_i=1$. In other words, a state
$|\Psi\rangle$
is physical only
if $D_i |\Psi\rangle =|\Psi\rangle$ for every $i$. In the Majorana
representation we have $\sigma^\alpha_i\sigma^\alpha_j
=\gamma_i^\alpha\gamma_j^\alpha\eta_i\eta_j=
-i\hat{u}_{ij}\eta_i\eta_j$, in which the link operators
$\hat{u}_{ij}=i\gamma_i^\alpha\gamma_j^\alpha$ mutually commute and
also commute with the Hamiltonian.
Since $\hat{u}_{ij}^2=1$, $\hat{u}_{ij}$ can be considered
as $c$-numbers with values $u_{ij}=\pm 1$, so that the Kitaev model
is equivalent to a free model of $\eta$ Majorana fermions coupled to
static $Z_2$ gauge fields
\cite{Kitaev06,Xiang07,Lee07,Yao07,Chen07,dusuel2008,nussinov2009}.
The ground state of such
a model is given by the direct product of a $Z_2$ gauge
configuration $\left|u\right\rangle$ and the corresponding Majorana
fermion ground state $\left|\phi(u)\right\rangle$. Here the
configuration $u$ is determined by minimizing the fermion ground
state energy. There is a macroscopic ground state degeneracy in the
enlarged Hilbert space, because each state
$\left|u\right\rangle\otimes\left|\phi(u)\right\rangle$ is
degenerate with all the states
$\left|u'\right\rangle\otimes\left|\phi(u')\right\rangle$ with $u'$
gauge equivalent to $u$. However, such a degeneracy is removed when
the constraint $D_i=1$ is applied. The physical ground state is the
``gauge'' average of the degenerate states, implemented by the
projection \cite{Kitaev06}:
\begin{eqnarray}
\ket{\Psi}
&=&\frac1{\sqrt{2^{N+1}}}\sum_{g} D_g \ket u\otimes\ket{\phi(u)}
\label{eq:proj}
\end{eqnarray}
where $N$ is the total number of lattice sites, $g$ denotes a set of
lattice sites, and $D_g=\prod_{i\in g} D_i$. We define
$D=\prod_{i\in {\cal L}} D_i$ with ${\cal L}$ the set of all lattice
sites. The sum $\sum_g$ is taken over
all possible subsets $g$ of ${\cal L}$. Note that, in Eq. (\ref{eq:proj}), we
implicitly assumed that $D\ket{u}\otimes\ket{\psi(u)} = \ket{u}\otimes\ket{\psi(u)}$
because states with $D=-1$ will be annihilated by the projection.
Consequently, we have $D_g=DD_{\bar{g}}$ for the complement $\bar{g}={\cal L}-g$, so that
$D_g\ket u\otimes\ket{\phi(u)}=D_{\bar g={\cal L}-g}\ket
u\otimes\ket{\phi(u)}$. In other words there are only $2^{N-1}$
inequivalent
gauge transformations, as expected.

\begin{figure}[b]
\includegraphics[scale=0.25]{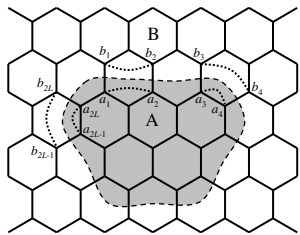}
\caption{The schematic
honeycomb lattice
is bipartitioned into two parts $A$
and $B$. The partition boundary (dashed line) cuts the links
$\overline{a_nb_n}$, $n=1,\cdots,2L$. New $Z_2$ gauge variables (see
text)
$\hat w_{A,n}$ and $\hat w_{B,n}$ are introduced on the new (dotted) links  $\overline{a_{2n-1}a_{2n}}$ and $\overline{b_{2n-1}b_{2n}}$, $n=1,\cdots,L$, respectively.
}
\label{fig:honeycomb}
\end{figure}

We define the Kitaev model on a torus and bipartite the lattice into
subsystems $A$ and $B$, as shown in Fig.
\ref{fig:honeycomb}. The EE between $A$ and $B$ is defined as
$S=-{\rm Tr}_A\left[\rho_A\log\rho_A\right]$, where $\rho_A=\Tr_B
\rho=\Tr_B\ket\Psi\bra\Psi$ is the reduced density matrix of $A$. To
calculate the EE, we will follow the ``{\it replica
trick}" introduced in Ref. \cite{Calabrese04}
\bea S=-\Tr_A \big[\rho_A \log \rho_A\big]
=-\frac{\partial}{\partial n} \Tr_A \big[\rho_A^n\big]
\Big|_{n=1}.\label{entropydef} \eea
The entanglement entropy
  can be
obtained if we can compute $\Tr_A\kd{\rho_A^n}$ for arbitrary
positive integer $n$ and then extrapolate the result to
$n\in\mathbb{R}$.

To obtain $\rho_A$, we trace out the spin degree of freedom in $B$, which normally can be carried out in terms of fermions and gauge fields. However, the gauge fields on the partition boundary are shared by $A$ and $B$; so we regroup those gauge fields on the boundary links to introduce new $Z_2$ gauge variables which lives in $A$ and $B$ exclusively, as shown in Fig. \ref{fig:honeycomb}. (see supplementary material \cite{SM} for details.) The calculation of $\textrm{Tr}[\rho_A^n]$ requires some careful treatment of the gauge transformation but is a well-defined mathematical procedure. Thus, we will leave the details involved in obtaining $\rho_A$ and $\textrm{Tr}\rho^n_A$ to the supplementary material [29] and present only the final result here:
\bea
\Tr_A[\rho^n_A]=\Tr_{A,G}[\rho^n_{A,G}]
\cdot\Tr_{A,F}[\rho^n_{A,F}],\label{DMequation} \eea
for any positive integer $n$. Here
$\rho_{A,F}=\Tr_B[\ket{\phi(u)}\bra{\phi(u)}]$ and $\rho_{A,G}=\Tr_{B}[\ket{G(u)}\bra{G(u)}]$ are the reduced
density matrices for the free Majorana fermion state $\ket{\phi(u)}$,
and a pure $Z_2$ gauge field \cite{Chamon07}, respectively, and the ground
state of the $Z_2$ gauge field $\ket{G(u)}$ is given by a equal
weight superposition of all the $2^{N-1}$ gauge field configurations
$\ket{\tilde{u}}$ that are gauge equivalent to $\ket{u}$, {\it
i.e.}, $\ket{G(u)}=2^{-(N-1)/2}\sum_{\tilde{u}\simeq u}\ket
{\tilde{u}}$. Physically, such a simplification occurs because the effect of the
gauge transformation $D_g$ on the fermion state in region $B$ is
canceled out once the trace over gauge field configurations is
taken.

Combining Eq. (\ref{DMequation}) and Eq. (\ref{entropydef}), it is
now obvious that the EE $S$ can be separated into
gauge field part $S_G$ and fermion part $S_F$ as follows: \bea
S=S_G+S_F. \label{entropysumrule}
\eea
Eq. (\ref{DMequation}) and (\ref{entropysumrule}) are among the
central results of this work. By explicit calculation \cite{SM} one can obtain $\Tr_{A,G}[\rho^n_{A,G}]=2^{-(L-1)(n-1)}$, so that  
$S_G=(L-1)\log 2$. As will be shown below, the fermion part has the
form $S_F=\alpha L+o(1)$, where $\alpha$ is a positive constant and
$o(1)$ represents terms which vanish as $L\to \infty$. In the
thermodynamic limit, the total entanglement entropy is given by \be
S=(\alpha+\log 2)L-\log 2, \ee from which we conclude that the TEE
is $S_{\mathrm{topo}}=-\log 2$. Our derivation is valid for all
phases of Kitaev model, including the Abelian ($Z_2$ gauge theory),
non-Abelian (Ising anyon)
phases, and also gapless phases. Thus our result
directly proves that the TEE for the Abelian and non-Abelian phases
are identical, as expected from the total quantum dimensions of
their quasiparticles \cite{Wang}.

\begin{figure}[b]
\includegraphics[scale=0.35]{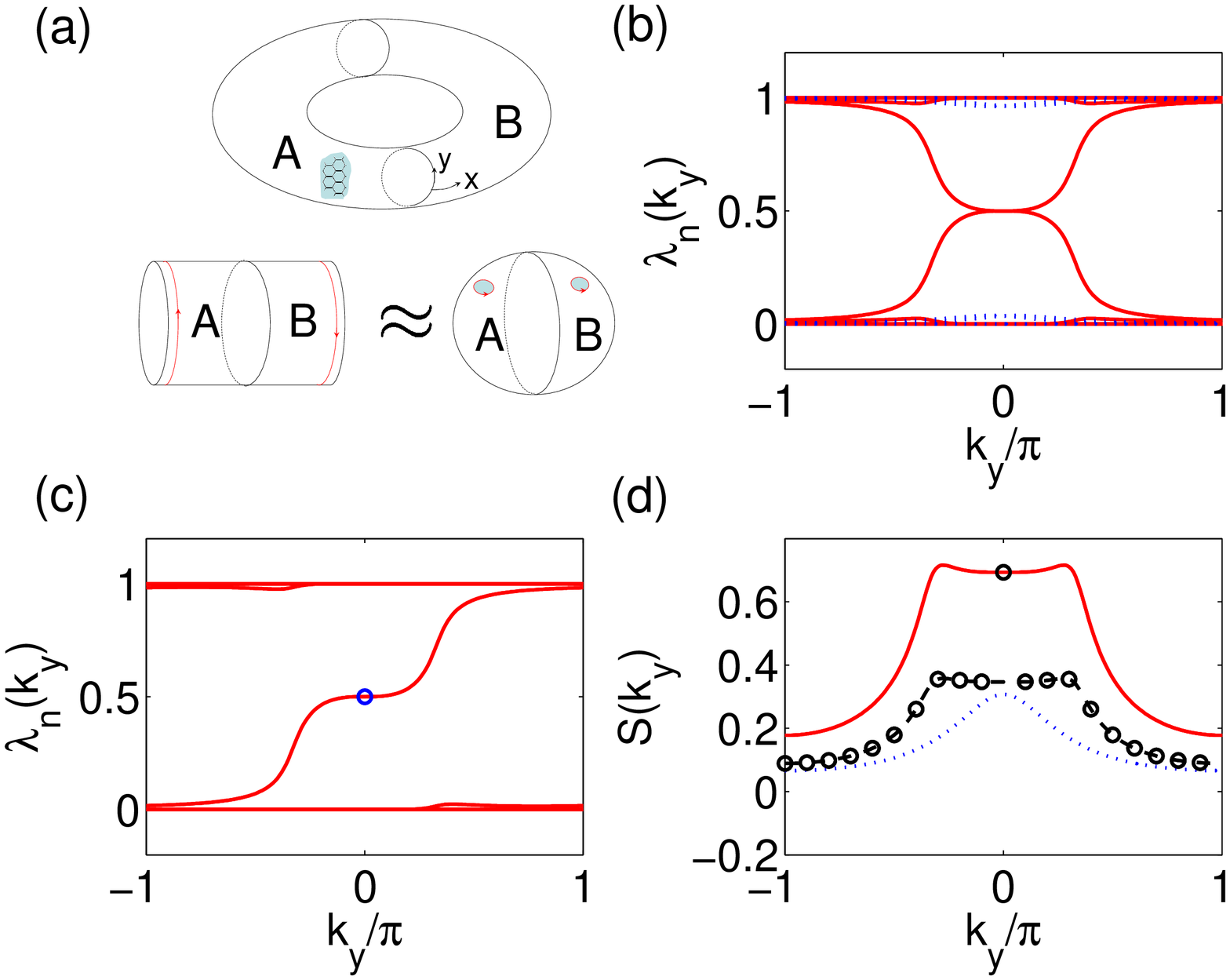}
\caption{(Color online) (a) Schematic picture of a torus and a
cylinder, each split to two regions $A$ and $B$. The cylinder is
equivalent to a sphere with two quasi-particles. (b) The
entanglement spectrum $\lambda_n(k_y)$ versus $k_y$ for non-Abelian
(red solid lines) and Abelian (blue dotted lines) state on torus.
Here and below, we take the parameters $J_x=J_y=J_z=1$ and
next-nearest neighbor coupling $J'=0.2$ for the non-Abelian
state, and $J_x=J_z=1,~J_y=2.5,~J'=0.2$ for the Abelian state.
(c) The entanglement spectrum for the non-Abelian state on cylinder.
The blue circle marks an additional state with $\lambda=1/2$ at
$k_y=0$. (d) The entropy $S(k_y)$ versus $k_y$ for non-Abelian (red
solid line) and Abelian (blue dotted line) states on the torus and
for non-Abelian state on cylinder (black dashed line with circles).
} \label{fig:fermion}
\end{figure}

Despite its trivial contribution to TEE, the fermion sector $S_F$ is
responsible for all the essential differences between the Abelian
and non-Abelian phase of Kitaev model in their quantum entanglement
properties. The EE of a free fermion system can be computed by the
method introduced in Ref. \cite{Peschel03}. To obtain an explicit
understanding to the fermion EE we consider a torus divided by two
parallel circles into $A$ and $B$ regions, as shown in Fig.
\ref{fig:fermion} (a). The boundary circle is along the $\hat{y}$
direction. On the torus, the free Majorana fermion Hamiltonian can
be block-diagonalized in the basis of $k_y$:
$H=\sum_{i,j}\eta_i\eta_jh_{ij}
=\sum_{x,x',k_y}\eta_{x}^\dagger(k_y)h_{xx'}(k_y) \eta_{x'}(k_y).$
Thus the system consists of decoupled one-dimensional
subsystems of each $k_y$.
The EE is given
by \cite{Peschel03}
\bea
S_F=-\frac12\sum_{n,k_y}\left[\lambda_n\log
\lambda_n+(1-\lambda_n)\log(1-\lambda_n)\right](k_y),
\label{torusEE}\eea
where $\lambda_n(k_y)$ are the eigenvalues of the
single-particle correlation function   $C_{xx'}(k_y)=\left\langle
\eta_x^\dagger(k_y)\eta_{x'}(k_y)\right\rangle$ for each $k_y$.
$\lambda_n$ plays the role of
Fermi-Dirac distribution
$1/(e^{\beta \epsilon_n}+1)$ in
thermal entropy, so that $\lambda_n=0~(1)$ corresponds to fully
occupied (unoccupied) states, respectively.
The ``entanglement spectrum'' (ES) $\lambda_n(k_y)$ has been computed numerically for both non-Abelian and Abelian
phases, as shown in Fig. \ref{fig:fermion} (b), for the Kitaev model
with three-spin terms $J'$ \cite{Lee07}.
The ES is gapped for the Abelian phase, and gapless for the
non-Abelian phase, similar to the edge states in the energy
spectrum. Similar observation has been made in topological
insulators and superconductors \cite{Ryu06, Ashvin, Lukasz,bray2009} and in
FQH systems \cite{Li07}. The two gapless
branches in the
ES come from the two boundaries
between $A$ and $B$. Since $\lambda_n(k_y)$'s are smooth functions
of $k_y$, we see from Eq. (\ref{torusEE}) that in the continuum
limit $S_F = \sum_{k_y}S(k_y) \simeq L \int S(k_y)\frac{dk_y}{2\pi}$ satisfies
the area law. It is interesting to note that a ``gap" always exists between the edge states and other bulk states with $\lambda_n(k_y)$ close to $0$ or $1$, which is analogous to the ``entanglement gap" studied in Ref. \cite{thomale2009b} for FQH system.

The situation becomes more interesting when we consider a cylinder with periodic boundary condition (PBC)
and the partition shown in
Fig. \ref{fig:fermion} (a). As shown in Fig. \ref{fig:fermion} (c),
in the non-Abelian phase the numerical calculation 
gives only one branch of ``gapless" states in the entanglement
spectrum. Physically, this is because the coupling through the other
boundary between $A$ and $B$ is removed by the open boundary
condition. However, at $k_y=0$ there is one {\em isolated}
additional state with $\lambda=1/2$, as shown by the blue circle in
Fig. \ref{fig:fermion} (c), which is due to the non-local
entanglement between the two Majorana zero modes at the open
boundary. Consequently, the entropy $S(k_y)$ is not a smooth
function of $k_y$ but has an additional $\log \sqrt{2}$ contributed
by $k_y=0$, as shown in Fig. \ref{fig:fermion} (d). Compared with
the torus case, in the thermodynamic limit we get $S_F=\alpha L+\log
\sqrt{2}$, which
shows explicitly that in the non-Abelian phase a cylinder with PBC
is topologically equivalent to a sphere with two non-Abelian
quasiparticles (usually named as $\sigma$ particles), as illustrated
in Fig. \ref{fig:fermion} (a). Each particle carries a
$\log\sqrt{2}$ entropy which is solely contributed by the fermion
sector.

Besides the
EE, more information
is contained in our result. The fact that Eq. (\ref{DMequation})
holds for any positive integer $n$ indicates that the many-body entanglement
spectrum---the eigenvalue spectrum  of $\rho_A$ is the direct
product of the ones of $\rho_{A,G}$ and $\rho_{A,F}$. From ${\rm
Tr}_{A,G}\left[\rho^n_{A,G}\right]=2^{-(L-1)(n-1)}$, one can know
that $\rho_{A,G}$ has $2^{L-1}$ nonzero eigenvalues, all of which
are degenerate and have the value of $2^{-(L-1)}$. Consequently all
non-vanishing eigenvalues of $\rho_A$ are given by those of the
Majorana fermion reduced density matrix $\rho_{A,F}$ times
$2^{-(L-1)}$. Thus the low ``energy'' ({\it i.e.}, close to the
maximal eigenvalue of $\rho_A$) feature in the entanglement spectra
of $\rho_A$  can be entirely characterized by its fermionic part,
which is gapped in the Abelian phase and gapless in the non-Abelian
phase, as shown in Fig. \ref{fig:fermion} (b).

\begin{figure}
\includegraphics[scale=0.4]{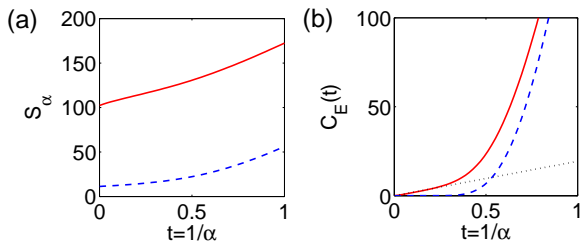}
\caption{(Color online) (a) Renyi entropy $S_\alpha$, and (b)
capacity of entanglement $C_E$ defined by Eq. (\ref{Ecapacity}), of
non-Abelian (red solid line) and Abelian (blue dashed line) states.
The black dotted line is a linear fitting. The parameters are the
same as those in Fig. 2.} \label{fig:renyi}
\end{figure}

Such a qualitative difference in the entanglement spectrum can be
characterized by the Renyi entropy \cite{Renyi} $
S_\alpha=\frac1{1-\alpha}\log{\rm Tr}\rho^\alpha$, which reduces to
the
EE (or von Neumann entropy) at $\alpha\to 1$.
According to Eq. (\ref{DMequation}) the Renyi
entropy of Kitaev model is given by
$S_\alpha=S_{F\alpha}+S_{G\alpha}$ for any $\alpha$, with
$S_{G\alpha}$ and $S_{F\alpha}$ the contribution from $Z_2$ gauge fields and fermions, respectively. From $\textrm{Tr}_{A,G}[\rho^n_{A,G}]=2^{-(L-1)(n-1)}$, one can see that $S_{G\alpha}=S_G=(L-1)\log 2$. Thus the TEE in $S_{G\alpha}$ is $\alpha$ independent, which is a generic property of the string-net models\cite{Levin2003,flammia2009}.
 The $\alpha$
dependence of $S_{F\alpha}$ in the Abelian and non-Abelian phases
has qualitative difference due to their different entanglement
spectra. If we define $\rho=e^{-\mathcal{H}}$, the quantity
$S_\alpha(1-1/\alpha)=-\frac1\alpha\log{\rm
Tr}e^{-\alpha\mathcal{H}}$ is the same as the free energy of a
thermal system with Hamiltonian $\mathcal{H}$ and temperature
$t=1/\alpha$. The behavior of the low energy spectrum of
$\mathcal{H}$ can be obtained from the following quantity:
\bea
C_E(t)=-t\frac{\partial^2}{\partial
t^2}\left[(1-t)S_{1/t}\right],\label{Ecapacity}
\eea
which is termed as ``{\it capacity of entanglement}" and is
the analog of heat capacity $C_v$ in a thermal system. The explicit
expression of $S_\alpha$ and $C_E(t)$ is given in the Supplementary
material, which leads to the numerical results shown in Fig.
\ref{fig:renyi}. As expected, in the limit of $t\rightarrow 0$,
$C_E(t)$ vanishes exponentially for Abelian phase but linearly for non-Abelian phase, since the latter
has a gapless
entanglement spectrum with constant density of state. More generically, if the entanglement Hamiltonian $\mathcal{H}$ describes a $(1+1)$ dimensional conformal field theory (CFT) in long wavelength limit \cite{KitaevPreskill,Li07} , the capacity of entanglement is given by $C_E(t)=(\pi cL/3v)t$ for $t\rightarrow 0$, with $L$ the length of the boundary, and $c$ and $v$ the central charge and velocity of the CFT, respectively \cite{affleck1986}. Moreover, if $\mathcal{H}$ describes a critical theory with dynamical exponent $z$, from dimensional analysis one can obtain the asymptotic behavior $C_E(t)\propto Lt^{1/z}$ for $t\rightarrow 0$. Thus we see that the capacity of entanglement characterizes some important qualitative behavior of the entanglement spectrum in generic systems.

We sincerely thank S.-B. Chung, J. Cui, E. Fradkin, T. L. Hughes, Y. Ran, S. Ryu, Z. Wang, and X.-G. Wen for helpful
discussions. This work is supported, in part, by DOE grants DE-AC02-05CH11231 (HY) at Berkeley and DF-FG02-06ER46287 (HY) and DE-AC02-76SF00515 (XLQ) at Stanford.

\begin{widetext}
\section{Supplementary material}
\section*{A: Derivation of Eq. (4)}
\renewcommand{\theequation}{A\arabic{equation}}  
\setcounter{equation}{0} 
To obtain $\rho_A$, we trace out the spin degrees of freedom in $B$. For the spins in $B$ away from the partition boundary, the trace can
be carried out in the Hilbert space of fermions and gauge field respectively. However, the spins on the boundary sites need more
careful treatment. Suppose the links across the boundary are denoted as $\overline{a_nb_n}$, $n=1,\cdots 2L$,
as shown in Fig. 1. (Note that we implicitly assumed the boundary length is even here.
The odd boundary length case need some extra care but our results
below remain valid.) The gauge field $\hat u_{a_nb_n}=i\gamma^\alpha_{a_n}\gamma^\alpha_{b_n}$ residing on
$\overline{a_nb_n}$
has spin degrees of freedom from both $A$ and $B$. To trace out $B$
we introduce new $Z_2$ variables
 $\hat w_{A,n}
=i\gamma^\alpha_{a_{2n-1}}\gamma^\beta_{a_{2n}}$ and $\hat w_{B,n} =i
\gamma^\alpha_{b_{2n-1}}\gamma^\beta_{b_{2n}}$,  which are defined on the dotted links in Fig. 1 and belong to either $A$ or $B$ exclusively.
 The eigenstates of
$\hat{w}_{A,n},~\hat{w}_{B,n}$ are related to those of
$\hat{u}_{a_{2n-1}b_{2n-1}},~\hat{u}_{a_{2n}b_{2n}}$ by a unitary
transformation.
Since $\ket u$ is a direct product of gauge fields, we denote $\ket u=\ket {u_A, u_B,u_p}$, where $\ket{u_A}$, $\ket{u_B}$, and $\ket {u_p}$
are gauge fields in $A$, $B$, and on the links across the partition boundary respectively.
In term of eigenstates of $\hat w_A$ and $\hat w_B$ defined above, we obtain
\bea
\ket {u_p}=\frac{1}{\sqrt{2^L}}\sum_{w_A=w_B=\{\pm 1\}}\ket
{w_A,w_B},
\label{basischange}
\eea
if all boundary links have
eigenvalues $u_{a_nb_n}=1$. For more generic values of $u_{a_nb_n}$,
the only change are the signs of the terms in the right hand side of Eq. (\ref{basischange}),
which does not enter the reduced density matrix we are interested in.  Here  $w_{A(B)}=\{w_{A(B),n}\}$ and $w_A=w_B$ means $w_{A,n}=w_{B,n}$ for all $n=1,...,L$.

In the new basis, the physical eigenstate
is given
by \bea \ket\Psi=\frac{1}{\sqrt{2^{N+L+1}}}\sum_{g,w_A=w_B} D_g
\ket{u_A,w_A;u_B,w_B}\ket{\phi(u)},\label{gsnewbasis} \eea from
which one can obtain the reduced density matrix $\rho_A$ and compute
${\rm Tr}\left[\rho_A^n\right]$. 

We define
$X_g=i^{|g|(|g|-1)/2}\prod_{j\in g}\gamma^x_j\gamma^y_j\gamma^z_j$
and $Y_g=i^{|g|(|g|-1)/2}\prod_{j\in g}\eta_j$, where $|g|$ denotes
to the number of sites in $g$ and the ordering of sites in the two
products is implicitly taken to be the same such that $X_gY_g=D_g$.
$X_g$ and $Y_g$ are the gauge transformation operators that only act
on the gauge fields and Majorana fermions, respectively. We further
denote $X_g=(-i)^{|g_A||g_B|}X_{g_B\equiv g\cap B}X_{g_A\equiv g\cap A}$ and
$Y_g=i^{|g_A||g_B|}Y_{g_A}Y_{g_B}$, where $X_{g_A}=i^{|g_A|(|g_A|-1)/2}\prod_{i\in
g_A}\gamma^x_i\gamma^y_i\gamma^z_i$ and
$Y_{g_A}=i^{|g_A|(|g_A|-1)/2}\prod_{i\in g_A}\eta_i$ are the gauge
transformation operator acting on $A$ region, and similar for
$X_{g_B}$ and $Y_{g_B}$. The ground state (5) is
written as \bea
\ket\Psi&=&\frac{1}{\sqrt{2^{N+L+1}}}\sum_{g,w_A=w_B} X_g
\ket{u_A,w_A;u_B,w_B}\cdot Y_g\ket{\phi(u)},\nn\\
&=& \frac{1}{\sqrt{2^{N+L+1}}}\sum_{g,w}  X_{g_B}\ket{u_B,w}\cdot X_{g_A}
\ket{u_A,w}\cdot
Y_{g_A}Y_{g_B}\ket{\phi(u)}.
\eea
In the last line we have denoted $w_A=w_B$ as $w$.
The reduced density matrix $\rho_A$ is expressed as
\bea
\rho_A&=&{\rm Tr_B}\big[\ket\Psi\bra\Psi\big]\nn\\
&=&\frac1{2^{N+L+1}} \sum_{g,g',w,w'}\bra{u_B,w'}X^\dag_{g'_B}X_{g_B} \ket{u_B,w} X_{g_A}
\ket{u_A,w}\bra{u_A,w'}X^\dag_{g'_A}
\cdot{\rm Tr}_{B,F}\left[Y_{g_A}Y_{g_B}\ket{\phi(u)}
\bra{\phi(u)}Y^\dag_{g'_B}Y^\dag_{g'_A}\right].~~~~~
\eea

To make the interproduct $\bra{u_B,w'} X^\dag_{g'_B}
X_{g_B}\ket{u_B,w}$ nonzero, the gauge transformation $X_{g_B}$ and
$X_{g'_B}$ must be either identical in B region, or different by a
gauge transformation on all sites in the B region. We define $X_B$
as $X_g$ with $g=B$, and similarly for $X_A$ and $Y_{A,B}$. Note
that $X^\dag_{g_B}=(-1)^{|g_B|} X_{g_B}$. So it is consistent to
define $B-g_B$ such that $X^\dag_{B- g_B}X_{g_B}=X_B$ for any $g_B$,
where $X_B\equiv X_{g=B}$. Consequently, we have \bea
\bra{u_B,w'}X^\dag_{g'_B}X_{g_B}\ket{u_B,w}
=\delta_{w,w'}\left(\delta_{g'_B,g_B}
+x_B(w)\delta_{g'_B,B-g_B}\right)\label{interproduct} \eea with \bea
x_B(w)=\bra{u_B,w}X^\dag_{B-g_B}X_{g_B}\ket{u_B,w}
=\bra{u_B,w}X_B\ket{u_B,w} =\prod_{\overline{ij}\in
B}u_{ij}\prod_{n=1}^Lw_{n}.\label{parityxB} \eea

Using this result $\rho_A$ can be simplified to
\bea
\rho_A&=&\frac1{2^{N+L+1}}\sum_{g_A,g'_A,w}\sum_{g_B}X_{g_A}
\ket{u_A,w}\bra{u_A,w}X^\dag_{g'_A}
\cdot
Y_{g_A}{\rm Tr}_{B,F}\left[\ket{\phi(u)}
\bra{\phi(u)}\left(1+Y_B\right)\right]Y^\dag_{g'_A},\nn\\
&=&\frac1{2^{N_A+L}}\sum_{g_A,g'_A,w}X_{g_A}
\ket{u_A,w}\bra{u_A,w}X^\dag_{g'_A}
\cdot
Y_{g_A}{\rm Tr}_{B,F}\left[\ket{\phi(u)}
\bra{\phi(u)}\left(\frac{1+x_B(w)Y_B}2\right)\right] Y^\dag_{g'_A}.
\eea
Notice that $Y_B=i^{|B|(|B|-1)/2}\prod_{j\in B}\eta_j$ is actually the fermion number parity operator in B region, we see that the operator $\frac{1+x_B(w)Y_B}2$ is actually the projector
of the fermion number parity $Y_B=x_B(w)$. For convenience, we define
\bea
P^{x_B(w)}_{B,F}=\frac{1+x_B(w)Y_B}2
\eea
and similar for A region. The reduced density matrix
\bea
{\rm Tr}_{B,F}\left[\ket{\phi(u)}
\bra{\phi(u)}P_{B,F}^{x_B(w)}\right]=\rho_{A,F}^{x_B(w)}
\eea
is the fermion density matrix with restriction to the fixed fermion number parity $x_B(w)$ in the B region. By using similar interproduct formula (\ref{interproduct})
for A region, $\rho_A^2$ can be computed as
\bea
\rho_A^2&=&\frac{1}{2^{N_A+2L-1}}\sum_{g_A,g'_A,w}X_{gA}
\ket{u_A,w}\bra{u_A,w}X^\dag_{g'A}\cdot
Y_{gA}\rho_{A,F}^{x_B(w)}P^{x_A(w)}_{A,F}\rho_{A,F}^{x_B(w)}Y_{g'A}
\eea
The same calculation can be repeated to obtain
\bea
\rho_A^n&=&\frac{1}{2^{N_A+nL-n+1}}\sum_{g_A,g'_A,w}X_{gA}
\ket{u_A,w}\bra{u_A,w}X_{g'A}\cdot
Y_{gA}\rho_{A,F}^{x_B(w)}\left(P^{x_A(w)}_{A,F}\rho_{A,F}^{x_B(w)}\right)^{n-1}Y_{g'A}
\eea

Finally we take the trace of $\rho_A^n$ by using the same Eq. (\ref{interproduct}) to obtain
\bea
{\rm Tr}_A\left[\rho_A^n\right]=\frac1{2^{n(L-1)}}\sum_{w}{\rm Tr}_A\left(P^{x_A(w)}_{A,F}\rho_{A,F}^{x_B(w)}\right)^n
\eea
For Eq. (\ref{parityxB}), we see that flipping the sign of one link $w_i$, for some integer $1\leq i\leq L$, changes the sign of both $x_A(w)$ and $x_B(w)$.
If we define $\prod_{\overline{ij}\in A(B)}u_{ij}=p_{A(B)}$, then $x_{A(B)}(w)=p_{A(B)}\prod_{n=1}^Lw_n$, and the
summation over all $2^L$ configurations of $w=\{w_i,i=1,2,...,L\}$ leads to
\bea
{\rm Tr}_A\left[\rho_A^n\right]=\frac1{2^{(n-1)(L-1)}}{\rm Tr}_A\left[\left(P^{p_A}_{A,F}\rho_{A,F}^{p_B}\right)^n+\left(P^{-p_A}_{A,F}\rho_{A,F}^{-p_B}\right)^n\right]
\eea

In the main text we have discussed that the ground state $\ket{u}\ket{\phi(u)}$ must satisfy the constraint $D\ket{u}\ket{\phi(u)}=\ket{u}\ket{\phi(u)}$. Since
$D=X_AX_BY_AY_B$ and $X_AX_B\ket{u}=p_Ap_B\ket{u}$, we obtain $Y_AY_B\ket{\phi(u)}=p_Ap_B\ket{\phi(u)}$. In other words, the total fermion parity is fixed, so that 
$P_{A,F}^{p_A}P_{A,F}^{-p_B}\ket{\phi(u)} =P_{A,F}^{-p_A}P_{A,F}^{p_B}\ket{\phi(u)}=0$. Thus we have
\bea
\rho_{A,F}^{p_B}P_{A,F}^{-p_A}={\rm Tr}_{B,F} \left[\ket{\phi(u)}\bra{\phi(u)}P_{B,F}^{p_B}\right] P_{A,F}^{-p_A} =0
\eea
so that
\bea
{\rm Tr}_A\left[\rho_A^n\right]&=&\frac1{2^{(n-1)(L-1)}}{\rm Tr}_A\left[\left(P^{p_A}_{A,F}\rho_{A,F}^{p_B}\right) +\left(P^{-p_A}_{A,F}\rho_{A,F}^{-p_B}\right)\right]^n,\nn\\
&=&\frac1{2^{(n-1)(L-1)}}{\rm Tr}_A\left[\rho_{A,F}^n\right],
\label{eq:A15}
\eea
where $\rho_{A,F}=\rho_{A,F}^{+}+\rho_{A,F}^{-}={\rm Tr}_B\left[\ket{\phi(u)}\bra{\phi(u)}\right]$ is the free fermion density matrix without fermion number parity constraint
in the B region. 

To understand this result more intuitively, we note that 
\bea 
\Tr_{A,G}[\rho^n_{A,G}]=\frac1{2^{(n-1)(L-1)}},
\eea
where $\rho_{A,G}=\Tr_{B}[\ket{G(u)}\bra{G(u)}]$ is the reduced density
matrix of a pure $Z_2$ gauge field, and the ground
state of the $Z_2$ gauge field $\ket{G(u)}$ is given by a equal
weight superposition of all the $2^{N-1}$ gauge field configurations
$\ket{\tilde{u}}$ that are gauge equivalent to $\ket{u}$, {\it
i.e.}, $\ket{G(u)}=2^{-(N-1)/2}\sum_{\tilde{u}\simeq u}\ket
{\tilde{u}}$. It follows that Eq. (\ref{eq:A15}) can be written as \bea
\Tr_A[\rho^n_A]=\Tr_{A,G}[\rho^n_{A,G}]
\cdot\Tr_{A,F}[\rho^n_{A,F}]. \eea 
Thus, Eq. (4) in the main text is proved. 

\section*{B: Renyi entropy and entanglement capacitance of Majorana fermions}
 \renewcommand{\theequation}{B\arabic{equation}}
  \setcounter{equation}{0}  

As shown in Ref. [31], the free fermion reduced density
matrix always have the form \bea
\rho=\exp\left[-\sum_n\epsilon_n\gamma_n^\dagger\gamma_n\right]/\Omega
\label{manybodyDM} \eea with
$\Omega=\prod_n\left(1+e^{-\epsilon_n}\right)$ the normalization
constant and $\gamma_n$ fermion annihilation operators in the
diagonal basis of the density matrix. For a system of Majorana
fermions $\eta_i$, define the equal time correlation function in the
ground state $C_{ij}=\left\langle \eta_i\eta_j\right\rangle/2$ for
$i,j$ restricted to the region $A$. The EE between two parts $A$ and
$B$ is then given by \bea S_F=-\frac12{\rm Tr}\left[C\log
C+(1-C)\log(1-C)\right]\label{fermionentropy}\eea with $C$ the
matrix with entries of $C_{ij}$. The ``single particle density
matrix" $C$ plays the role of $\left(1+\te^{\beta h}\right)^{-1}$ in
the thermal entropy, with $h$ the single particle Hamiltonian and
the inverse temperature $\beta=1/T=1$.
The factor $1/2$ comes from the
fact that a Majorana fermion has half the degree of freedom of a
Dirac fermion.

The
eigenvalues $\lambda_n$ of the single particle density matrix $C$ in
Eq. (\ref{fermionentropy}) and
$\epsilon_n$ in Eq. (\ref{manybodyDM}) are related as \bea
\lambda_n=\frac{1}{e^{\epsilon_n}+1}\Rightarrow
e^{-\epsilon_n}=\frac{\lambda_n}{1-\lambda_n}.
\eea
Note that for Majorana fermion only $\epsilon_n\geq 0$ states are
summed over in Eq. (\ref{manybodyDM}). Using Eq. (\ref{manybodyDM})
we have \bea {\rm
Tr}\rho^\alpha=\frac1{\Omega^\alpha}\prod_{\epsilon_n\geq
0}\left(1+e^{-\alpha\epsilon_n}\right)\eea
so that
\bea
S_\alpha&=&\frac12\frac1{1-\alpha} \sum_n\left[\log\left(1+e^{-\alpha
\epsilon_n}\right) -\alpha\log\left(1+e^{-\epsilon_n}\right)\right],\nonumber\\
&=&\frac12\frac1{1-\alpha} \sum_n\log\left[\lambda_n^\alpha +\left(1-\lambda_n\right)^\alpha\right].
\eea

The ``capacity of entanglement" $C_E$ defined in Eq. (11)
can be obtained by 
\bea C_E&=&\frac t2\frac{\partial^2}{\partial t^2} \left[t\sum_n\log\left[\lambda_n^{1/t} +\left(1-\lambda_n\right)^{1/t}\right]\right],\nonumber\\
&=&\frac1{2t^2}\sum_n{\left( \frac{\epsilon_n}{2\cosh\left({\epsilon_n}/{2t}\right)} \right)^2}.
\eea
It should be noticed that all the results above are for free Majorana fermions. For free complex fermions the only difference is an additional
factor of $2$.

In the torus case we studied, $k_y$ is a good quantum number, and we
have 
\bea C_E=\frac{L}{2t^2}\sum_n\int
\frac{dk_y}{2\pi}{\left(\frac{\epsilon_n(k_y)} {2\cosh\left({\epsilon_n(k_y)}/{2t}\right)}\right)^2}.
\eea
For the Abelian phase $\epsilon_n$ has a gap since $\lambda_n(k_y)$
does not cross $1/2$. If the gap is $E_g={\rm
min}(|\epsilon_n(k_y)|)$, the asymptotic behavior of $C_E$ at
$t\rightarrow 0$ is $C_E\simeq e^{-E_g/t}/t^2$. For the non-Abelian
phase, as shown in Fig. 2(b) there is a gapless
branch of $\lambda_n(k_y)$ crossing $1/2$, which corresponds to
$\epsilon_n(k_y)$ crossing $0$. Near $k_y=0$ the asymptotic behavior
of $|\epsilon_n(k_y)|$ is $|\epsilon_n(k_y)|\simeq v|k_y|$, which
gives $C_E\propto t/v$. In the same way as the heat capacity in a
thermodynamic system, $C_E/t$ is constant at $t\rightarrow 0$ limit
and is proportional to the density of state $1/v$ at ``maximally
entangled limit" $\epsilon_n\rightarrow 0$ or $\lambda_n\rightarrow
1/2$.
\end{widetext}

\end{document}